\pdfoutput=1
\documentclass[12pt,a4paper,final]{iopart}
\usepackage{blindtext}
\usepackage{iopams}  
\usepackage{cite}
\usepackage{graphicx}
\usepackage{wrapfig}
\usepackage[breaklinks=true,colorlinks=true,linkcolor=blue,urlcolor=blue,citecolor=blue]{hyperref}
\usepackage{xcolor}
\usepackage{multicol}
\usepackage{bm}
\expandafter\let\csname equation*\endcsname\relax
\expandafter\let\csname endequation*\endcsname\relax
\usepackage{amsmath}
\newcommand{\abs}[1]{\lvert #1 \rvert}

\newcommand{\bigabs}[1]{\big\lvert#1\big\rvert}

\newcommand{\ord}[1]{\mathcal{O}(#1)}

\newcommand{\pd}[2]{\frac{\partial #1}{\partial #2}}
\newcommand{\braced}[2]{\lbrace#1,#2\rbrace}
\newcommand{\pdshort}[2]{{#1}^{\polangle}}

\renewcommand{\d}{\mathrm{d}}

\newcommand{\code}[1]{\texttt{#1}}
\newcommand{\gstwo}{\code{GS2}}
\newcommand{\gkv}{\code{GKV}}
\newcommand{\shat}{\hat{s}}

\newcommand{\dashsixty}{-60}
\newcommand{\jtsa}{JT\dashsixty{SA}}

\newcommand{\atantermdef}{\arctan\bigg( \frac{\tan\polangle}{\kappa} \bigg)}
\newcommand{\atanterm}{\vartheta}
\newcommand{\lnterm}{\frac{1}{2\elo\sqrt{\elo^2-1}} \log\Bigg(\frac{1+\sqrt{1 - \elo^{-2}}\sin\polangle}{1-\sqrt{1 - \elo^{-2}}\sin\polangle}\Bigg) }

\newcommand{\specone}{s}

\newcommand{\reference}{\text{ref}}

\newcommand{\nonadflucs}{h_\specone}

\newcommand{\feqlords}{F_{0,\specone}}

\newcommand{\Beq}{\bm{B}}
\newcommand{\gradmagBeq}{\nabla\magBeq}

\newcommand{\beq}{\bm{b}}
\newcommand{\magBeq}{B}
\newcommand{\polcurfunc}{I}
\newcommand{\Bfluc}{\delta\bm{B}}
\newcommand{\Bflucpar}{\delta{B}_\parallel}

\newcommand{\Eeq}{\bm{E}}

\newcommand{\Aflucpar}{\delta{A}_\parallel}
\newcommand{\gyropot}{\chi}\newcommand{\pot}{\phi}
\newcommand{\flucpot}{\delta\pot}

\newcommand{\Bref}{B_{ref}}

\newcommand{\kperp}{\bm{k}_\perp}
\newcommand{\magkperp}{\abs{\kperp}}

\newcommand{\ky}{k_y}

\newcommand{\kx}{k_x}

\newcommand{\freqlarm}[1]{\Omega_{#1}}
\newcommand{\freqlarms}{\freqlarm{\specone}}

\newcommand{\growthrate}{\gamma}
\newcommand{\realfreq}{\omega}

\newcommand{\larm}[1]{\rho_{#1}}

\newcommand{\larmi}{\larm{i}}
\newcommand{\larmref}{\larm{\reference}}
\newcommand{\macrolength}{a}

\renewcommand{\time}{t}
\newcommand{\pos}{\bm{r}}
\newcommand{\vpos}{\bm{v}}
\newcommand{\vposperp}{\vpos_\perp}
\newcommand{\magvposperp}{\abs{\vposperp}}

\newcommand{\polflux}{\psi}

\newcommand{\gradpolflux}{\nabla\polflux}
\newcommand{\minrad}{r}
\newcommand{\gradminrad}{\nabla\minrad}
\newcommand{\majrad}{R}
\newcommand{\majradzero}{\majrad_{0}}

\newcommand{\normtorflux}{\rho}

\newcommand{\torangle}{\zeta}
\newcommand{\polangle}{\theta}
\newcommand{\heightZ}{Z}
\newcommand{\clebscha}{\tilde{\alpha}}
\newcommand{\varpolangle}{\vartheta}
\newcommand{\gradtorangle}{\nabla\torangle}
\newcommand{\gradpolangle}{\nabla\polangle}

\newcommand{\gradclebscha}{\nabla\clebscha}
\newcommand{\jac}{\mathcal{J}}

\newcommand{\jacr}{\mathcal{J}_\minrad}
\newcommand{\jacprim}{J}

\newcommand{\vth}[1]{v_{\text{th},#1}}

\newcommand{\vthref}{\vth{\reference}}

\newcommand{\vpar}{v_\parallel}

\newcommand{\lightspeed}{c}
\newcommand{\vdrifts}{\bm{V}_{D,\specone}}
\newcommand{\exb}{\Eeq\times\Beq}
\newcommand{\vflucpot}{\bm{V}_\gyropot}

\newcommand{\zs}{Z_s}

\newcommand{\elcharge}{e}

\newcommand{\dens}[1]{n_{#1}}
\newcommand{\denss}{\dens{\specone}}

\newcommand{\plasmacurrent}{I_{p}}

\newcommand{\temp}[1]{T_{#1}}
\newcommand{\temps}{\temp{\specone}}
\newcommand{\tempi}{\temp{i}}
\newcommand{\tempe}{\temp{e}}

\newcommand{\pressure}{p}

\newcommand{\normpprim}{(\log\pressure)'}

\newcommand{\bess}{J}
\newcommand{\bessz}{\bess_0}
\newcommand{\bessf}{\bess_1}
\newcommand{\bessargs}{u_\specone}
\newcommand{\dummy}[1]{\hat{#1}}

\newcommand{\tri}{\delta}
\newcommand{\trir}{\tri'}
\newcommand{\elo}{\kappa}
\newcommand{\elor}{\elo'}
\newcommand{\locsafety}{\tilde{\safety}}
\newcommand{\safety}{q}
\newcommand{\locshear}{\tilde{s}}
\newcommand{\betaprim}{\beta'}

\newcommand{\invasprat}{\epsilon}
\newcommand{\shift}{\Delta}

\newcommand{\appropto}{\mathrel{\vcenter{
			\offinterlineskip\halign{\hfil$##$\cr
				\propto\cr\noalign{\kern1pt}\sim\cr\noalign{\kern-2pt}}}}}
\usepackage[toc,page]{appendix}
\begin{document}

\title[Impact of plasma shaping on tokamak microstability]{Impact of plasma shaping on tokamak microstability}

\author{O. Beeke$^{1}$, M. Barnes$^{1}$, M. Romanelli$^{2}$, M. Nakata$^3$, M.~Yoshida$^4$}
\address{$^1$Rudolf Peierls Centre for Theoretical Physics, University of Oxford, OX1 3PU, United Kingdom\\
$^2$Culham Centre for Fusion Energy, Culham Science Centre, Abingdon, Oxfordshire, OX14 3DB, United Kingdom\\
$^3$National Institute for Fusion Science, Toki 509-5292, Japan\\
$^4$National Institutes for Quantum and Radiological Science and Technology, Naka, Japan}
\ead{oliver.beeke@physics.ox.ac.uk}

\begin{abstract}
We have used the local-$\delta f$ gyrokinetic code \gstwo\ to perform studies of the effect of flux-surface shaping on two highly-shaped, low- and high- $\beta$ \jtsa-relevant equilibria, including a successful benchmark with the \gkv\ code. We find a novel destabilization of electrostatic fluctuations with increased elongation for plasma with a strongly peaked pressure profile. We explain the results as a competition between the local magnetic shear and finite-Larmor-radius (FLR) stabilization. Electromagnetic studies indicate that kinetic ballooning modes are stabilized by increased shaping due to an increased sensitivity to FLR effects, relative to the ion-temperature-gradient instability. Nevertheless, at high enough $\beta$, increased elongation degrades the local magnetic shear stabilization that enables access to the region of ballooning second-stability. 

\end{abstract}

\vspace{2pc}

\section{Introduction}

\begin{figure}
	\centering
	\def\svgwidth{0.4\linewidth}
\begingroup%
  \makeatletter%
  \providecommand\color[2][]{%
    \errmessage{(Inkscape) Color is used for the text in Inkscape, but the package 'color.sty' is not loaded}%
    \renewcommand\color[2][]{}%
  }%
  \providecommand\transparent[1]{%
    \errmessage{(Inkscape) Transparency is used (non-zero) for the text in Inkscape, but the package 'transparent.sty' is not loaded}%
    \renewcommand\transparent[1]{}%
  }%
  \providecommand\rotatebox[2]{#2}%
  \newcommand*\fsize{\dimexpr\f@size pt\relax}%
  \newcommand*\lineheight[1]{\fontsize{\fsize}{#1\fsize}\selectfont}%
  \ifx\svgwidth\undefined%
    \setlength{\unitlength}{375bp}%
    \ifx\svgscale\undefined%
      \relax%
    \else%
      \setlength{\unitlength}{\unitlength * \real{\svgscale}}%
    \fi%
  \else%
    \setlength{\unitlength}{\svgwidth}%
  \fi%
  \global\let\svgwidth\undefined%
  \global\let\svgscale\undefined%
  \makeatother%
  \begin{picture}(1,1)%
    \lineheight{1}%
    \setlength\tabcolsep{0pt}%
    \put(0,0){\includegraphics[width=\unitlength,page=1]{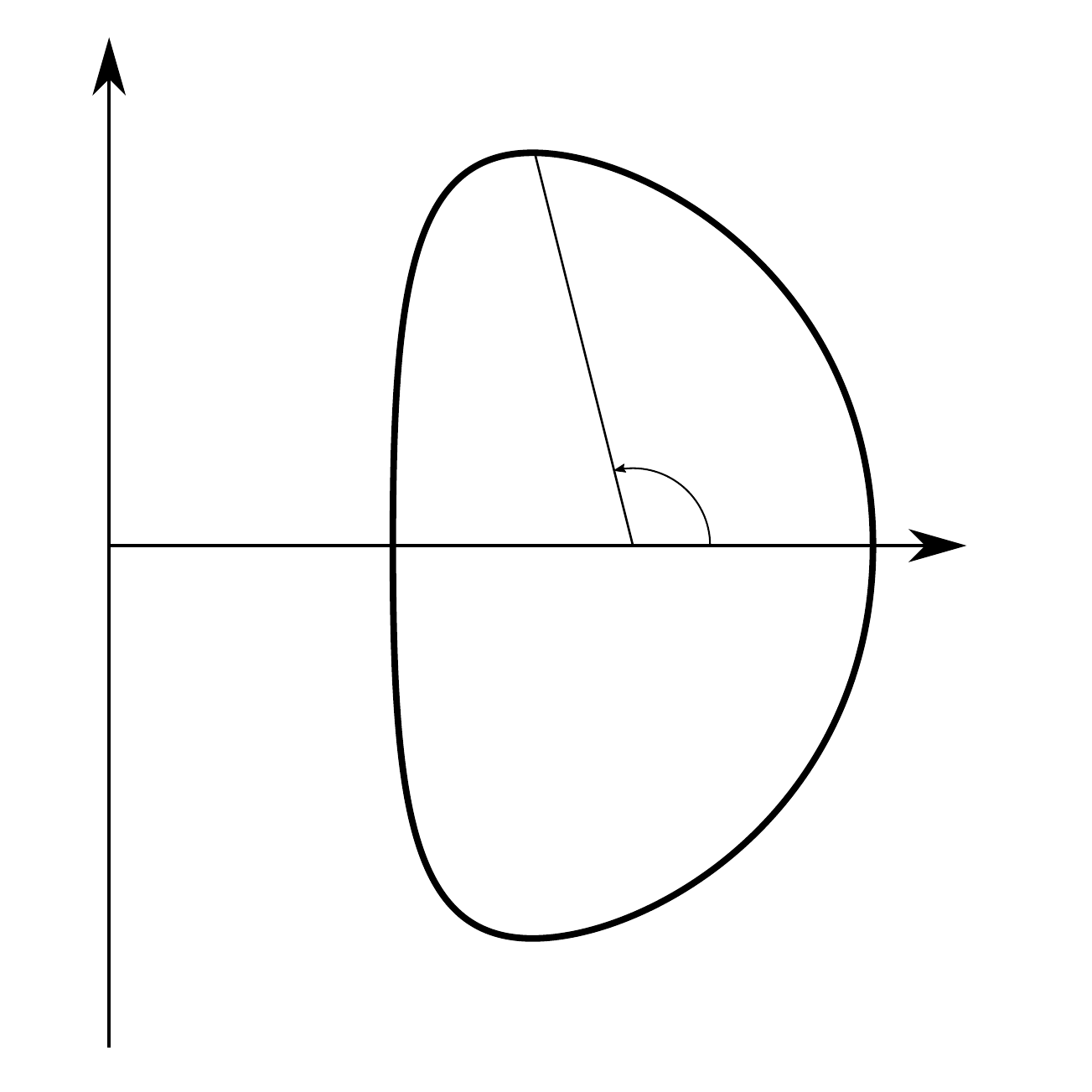}}%
    \put(0.6190541,0.56535754){\color[rgb]{0,0,0}\makebox(0,0)[lt]{\lineheight{1.25}\smash{\begin{tabular}[t]{l}$\polangle_{\rm{max}}$\end{tabular}}}}%
    \put(0.03783183,0.95334528){\color[rgb]{0,0,0}\makebox(0,0)[lt]{\lineheight{1.25}\smash{\begin{tabular}[t]{l}$\heightZ$\end{tabular}}}}%
    \put(0.8710044,0.4334013){\color[rgb]{0,0,0}\makebox(0,0)[lt]{\lineheight{1.25}\smash{\begin{tabular}[t]{l}$\majrad$\end{tabular}}}}%
    \put(0.27478566,0.02292843){\color[rgb]{0,0,0}\makebox(0,0)[lt]{\lineheight{1.25}\smash{\begin{tabular}[t]{l}$\majradzero(\minrad)$\end{tabular}}}}%
    \put(0,0){\includegraphics[width=\unitlength,page=2]{fs.pdf}}%
    \put(0.55300008,0.93614363){\color[rgb]{0,0,0}\makebox(0,0)[lt]{\lineheight{1.25}\smash{\begin{tabular}[t]{l}$2\minrad$\end{tabular}}}}%
    \put(0,0){\includegraphics[width=\unitlength,page=3]{fs.pdf}}%
    \put(0.03857143,0.46428574){\color[rgb]{0,0,0}\makebox(0,0)[lt]{\lineheight{1.25}\smash{\begin{tabular}[t]{l}$0$\end{tabular}}}}%
  \end{picture}%
\endgroup%

	\caption{\label{fig:fluxsurface}A typical flux-surface shape, with $\minrad$, $\majradzero$ and $\polangle_{\mathrm{max}}$ labelled. We define the midplane as the horizontal plane that intersects the widest point of the flux-surface; $\polangle=0$ coincides with the midplane on the outboard side.}
\end{figure}

One of the performance-limiting factors on tokamak experiments is the significant turbulence-dominated radial transport of heat, momentum and particles from the core plasma out to the edge. One method of affecting this turbulent transport is to change the shape of the axisymmetric flux-surfaces traced out by the confining magnetic field; this has been validated both experimentally and in numerical simulations. Flux-surface shape is often characterised by the elongation $\elo$ and triangularity $\tri$, where $\elo(\minrad)\equiv\heightZ(r,\polangle_\mathrm{max})/\minrad$ and $\tri(r)\equiv[\majrad_{0}(\minrad)-\majrad(\minrad,\polangle_\mathrm{max})]/\minrad$, where $\majrad(\minrad,\polangle)$ is the plasma major radius, $\polangle$ is the poloidal angle, $\heightZ(\minrad,\polangle)$ is the vertical distance above the midplane, $\minrad\equiv[R(r,0)-R(r,\pi)]/2$ is the plasma half-diameter at the flux-surface midplane, $\majradzero(\minrad)\equiv[R(r,0)+R(r,\pi)]/2$ is the major radius of the flux-surface center and $\polangle_\mathrm{max}$ is the poloidal angle of the maximum $Z$. A labelled example flux-surface shape is shown in Figure \ref{fig:fluxsurface}. The above expressions are valid for the up-down symmetric plasmas studied in this work, although similar expressions can be used to account for up-down asymmetry. Plasma shaping is known to affect the stability of magnetohydrodynamic (MHD) modes. Specifically, increased elongation and triangularity increase the threshold ``Troyon" plasma $\beta$ above which dangerous (e.g. external kink-ballooning) MHD instabilities are triggered: $\beta_{Troyon}=\beta_{N}\plasmacurrent/(a\Bref)$, where $\beta_{N}\simeq2\%$ is an empirically calculated scaling factor, $\macrolength$ is the minor radius of the last closed flux-surface (LCFS) and $\Bref$ is the reference magnetic field, defined as the on-axis toroidal field \cite{troyon1984}. The Troyon beta limit increases linearly with plasma current $\plasmacurrent$ which, for fixed safety factor $\safety$, is increased by elongation and triangularity. Shaping can however have other important consequences, particularly for vertical displacement events (VDE) which are associated with disruptions. Whilst increased elongation can ameliorate the aforementioned MHD instabilities, more severe controls for VDE are then required. This necessitates the careful investigation of shaping effects from both MHD/disruption and transport perspectives.
\paragraph{}The impact of shaping on transport due to micro-instabilities has been less-thoroughly studied, and those studies that have been performed have focused only on a relatively small region of the vast multi-dimensional shaping parameter space. Despite this, increased elongation is generally thought to be stabilizing \cite{kinsey2007,belli2008,angelino2009,ball2018uda}, whereas triangularity can have varying effects. Both the TCV and DIII-D experiments have reported that negative triangularity yields improved performance, allowing performance comparable to H-mode whilst maintaining L-mode edge profiles \cite{weisen1997,moret1997,camenen2007,marinoni2019}. Meanwhile, \cite{belli2008} has reported a dependence on elongation of the effect of triangularity, being stabilizing at large $\elo$ and somewhat destabilizing at moderate $\elo$. There are various explanations for these observed effects on the turbulence; some suggest that flux-surface shaping modifies locally the driving gradients \cite{weisen1997,moret1997}, whilst others point to its effect on local magnetic shear\cite{kendl2006}.

\paragraph{}In this work we extend an earlier modelling study by Nakata \cite{nakata2014} studying two JT-60SA-relevant magnetic equilibria \cite{giruzzi2020}. Following a benchmark between the local, $\delta f$-gyrokinetic codes \gkv\ \cite{watanabe2006,nakata2015} and \gstwo, we present a study on the effects of triangularity and elongation in these two equilibria, both with and without electromagnetic effects. In Section \ref{sec:model}, we present how shaping influences turbulence in the gyrokinetic framework, as a combination of the local-magnetic-shear stabilization of the toroidal drive, as well as finite-Larmor-radius (FLR) stabilization. In Section \ref{sec:results}, we present the results of the benchmark, followed by electrostatic linear and non-linear shaping scans in $\lbrace\tri,\elo\rbrace$. We use our model to explain the qualitative trends observed in these scans, including the novel result that increased elongation can be destabilizing. Finally, we present linear electromagnetic shaping scans, showing how kinetic-ballooning-modes (KBMs) react differently to flux-surface shape. The work is then summarized in Section \ref{sec:summary}.

\section{Model}\label{sec:model}
\paragraph{}To understand how plasma shaping affects the local, $\delta f$ gyrokinetic framework in which \gstwo\ works, we write the Fourier-analysed (perpendicular to the equilibrium magnetic field) collisionless gyrokinetic equation and field equations as:
\begin{equation}
\pd{\nonadflucs}{\time} + \big[\vpar\beq\cdot\nabla + i\kperp\cdot\vdrifts\big]\nonadflucs + \braced{\gyropot}{\nonadflucs} =  \bigg[\frac{\zs\elcharge \feqlords}{\temps}\pd{\gyropot}{\time} - \vflucpot\cdot\nabla\feqlords\bigg];
\end{equation}
\begin{equation}
0 = \sum_\specone\zs\elcharge\int d^3 \vpos\bigg(\nonadflucs \bessz(\bessargs) - \frac{\zs\elcharge\delta\phi}{\temps}\feqlords(\pos)\bigg);
\end{equation}
\begin{equation}
\magkperp^2\Aflucpar = \frac{4\pi}{\lightspeed}\sum_{s}\zs\elcharge\int d^3\vpos \hspace{1mm} \vpar \nonadflucs\bessz(\bessargs);\label{eq:parcureq}
\end{equation}
\begin{equation}
	\magkperp\Bflucpar = -\frac{4\pi}{\lightspeed}\sum_\specone\zs\elcharge\int\d^3\vpos \magvposperp\nonadflucs\bessf(\bessargs),\label{eq:perpcureq}
\end{equation}
where $\nonadflucs$ is the non-adiabatic part of the fluctuating distribution function for a species labelled $\specone$ (note that we have omitted the wavenumber index on fluctuating quantities to simplify notation), $\vpos$ is the particle velocity, $\bess_n$ is the $n^\textrm{th}$-order Bessel function of the first kind, $\zs\elcharge$ is the charge, $\temps$ is the temperature, $\feqlords$ is the assumed-Maxwellian equilibrium distribution function,
\begin{equation}
	\gyropot\equiv\bessz(\bessargs)(\flucpot-\vpar\Aflucpar) + \bessf(\bessargs)\frac{\magvposperp}{\magkperp}\Bflucpar
\end{equation} is the gyro-averaged gyrokinetic potential, $\flucpot$ is the fluctuating electrostatic potential, $\Aflucpar$ and $\Bflucpar$ are the parallel components of the magnetic vector potential and field, respectively, $\lbrace\dots\rbrace$ indicates the Fourier-transformed Poisson bracket, 
\begin{equation}
\vdrifts\equiv\frac{\beq}{\freqlarms}\times\bigg[ \bigg( \vpar^2 + \frac{\vposperp^2}{2}\bigg)\frac{\gradmagBeq}{\magBeq} + \vpar^2\frac{4\pi\pressure'}{\magBeq^2}\gradminrad  \bigg]
\end{equation} 
contains the non-$\exb$ magnetic drifts, ' indicates an $\minrad$-derivative, $\pressure$ is the pressure, $\freqlarms$ is the Larmor frequency, $\vflucpot\equiv i (\lightspeed\gyropot/\magBeq^2)\Beq\times\kperp$ is the drift velocity due to the interaction between the equilibrium magnetic field and the fluctuating electric and magnetic potentials, $\bessargs\equiv\abs{\kperp}\abs{\vposperp}/\freqlarms$ and $\kperp\equiv k_{\polflux}\gradpolflux + k_{\clebscha}\gradclebscha$ where $\polflux$ is the poloidal magnetic flux, used as a radial coordinate, and $\clebscha$ is the field-line label. 

\begin{figure}
	\centering
	\def\svgwidth{0.7\linewidth}
\begingroup%
  \makeatletter%
  \providecommand\color[2][]{%
    \errmessage{(Inkscape) Color is used for the text in Inkscape, but the package 'color.sty' is not loaded}%
    \renewcommand\color[2][]{}%
  }%
  \providecommand\transparent[1]{%
    \errmessage{(Inkscape) Transparency is used (non-zero) for the text in Inkscape, but the package 'transparent.sty' is not loaded}%
    \renewcommand\transparent[1]{}%
  }%
  \providecommand\rotatebox[2]{#2}%
  \newcommand*\fsize{\dimexpr\f@size pt\relax}%
  \newcommand*\lineheight[1]{\fontsize{\fsize}{#1\fsize}\selectfont}%
  \ifx\svgwidth\undefined%
    \setlength{\unitlength}{576bp}%
    \ifx\svgscale\undefined%
      \relax%
    \else%
      \setlength{\unitlength}{\unitlength * \real{\svgscale}}%
    \fi%
  \else%
    \setlength{\unitlength}{\svgwidth}%
  \fi%
  \global\let\svgwidth\undefined%
  \global\let\svgscale\undefined%
  \makeatother%
  \begin{picture}(1,0.375)%
    \lineheight{1}%
    \setlength\tabcolsep{0pt}%
    \put(0,0){\includegraphics[width=\unitlength,page=1]{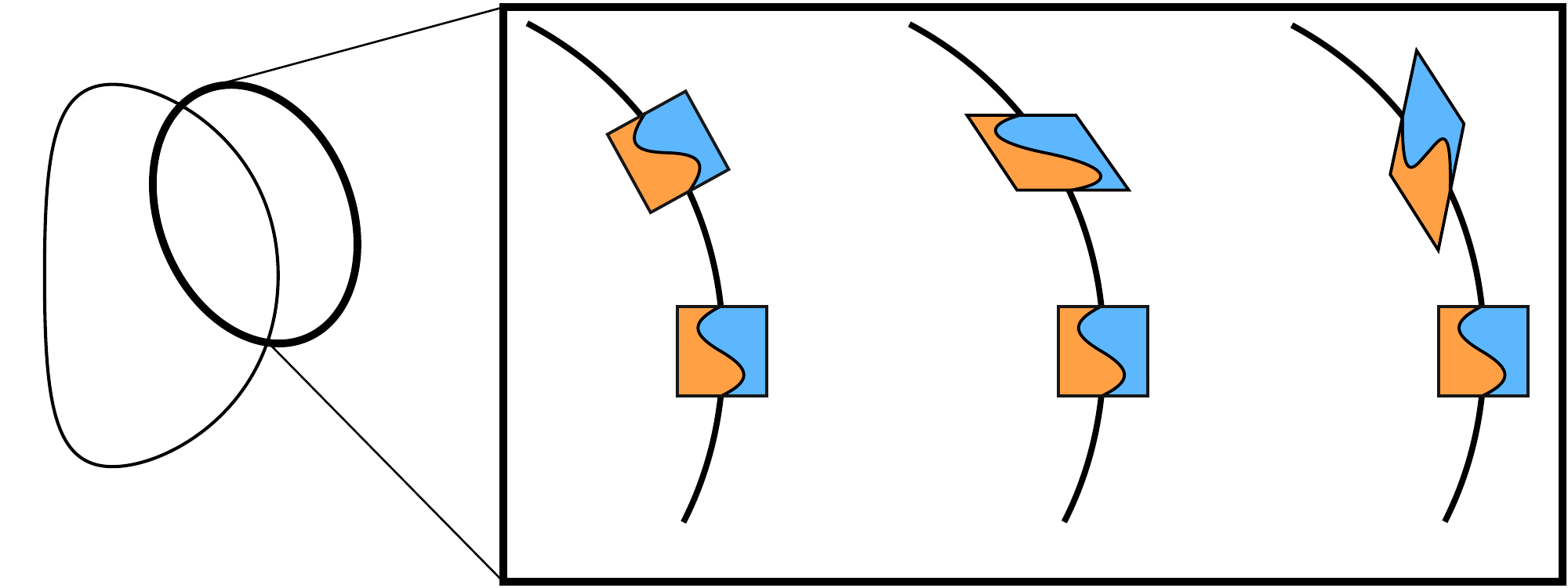}}%
    \put(0.4070542,0.01581137){\color[rgb]{0,0,0}\makebox(0,0)[lt]{\lineheight{1.25}\smash{\begin{tabular}[t]{l}$\hat{s}=0$\end{tabular}}}}%
    \put(0.64874096,0.01581137){\color[rgb]{0,0,0}\makebox(0,0)[lt]{\lineheight{1.25}\smash{\begin{tabular}[t]{l}$\hat{s}\simeq1$\end{tabular}}}}%
    \put(0.8928365,0.01581137){\color[rgb]{0,0,0}\makebox(0,0)[lt]{\lineheight{1.25}\smash{\begin{tabular}[t]{l}$\hat{s}<0$\end{tabular}}}}%
  \end{picture}%
\endgroup%

	\caption{\label{fig:locshear}The effect of magnetic shear on a ballooning-type perturbation. The instability is driven strongly at the outboard midplane where $\vdrifts\cdot\kperp$ is maximal. The perturbations are advected along the field lines to a different poloidal (and toroidal) position. With moderate positive magnetic shear, the outboard side lags behind the inboard side poloidally, so the perturbation retains some major-radial extent and remains strongly driven. For negative magnetic shear, the inboard side lags and the perturbation gains more vertical extent, and thus couples more weakly to the predominantly vertical magnetic drifts. A similar argument is used to explain why \textit{large} positive values of magnetic shear have a stabilizing effect.} 
\end{figure}
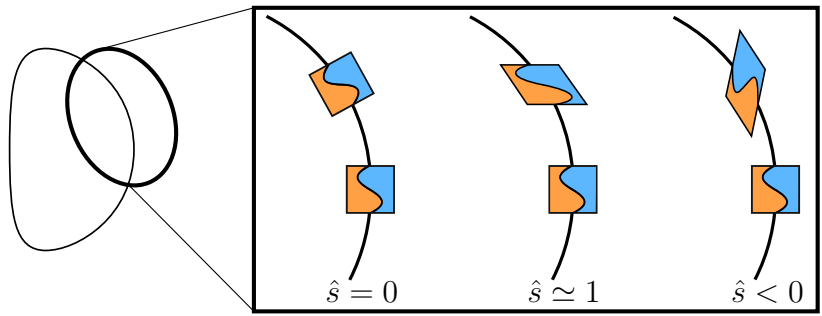

\paragraph{}The flux-surface shape enters these equations via $\gradpolflux$ and $\gradclebscha$ in $\kperp$. The perpendicular wavevector $\kperp$ can strengthen (weaken) the toroidal drive term if (mis)aligned with $\vdrifts$; this is closely linked to local-shear stabilization as shown in Figure \ref{fig:locshear}. It can also provide FLR stabilization via the gyro-average, manifested as the Bessel functions. To calculate $\gradpolflux$ and $\gradclebscha$, we work in toroidal $\lbrace\minrad,\polangle,\torangle\rbrace$ and cylindrical $\lbrace\majrad,\torangle,\heightZ\rbrace$ coordinates, where $\minrad$, $\polangle$, $\heightZ$ and $\majrad$ retain their previous definitions and $\torangle$ is the toroidal angle. By using the Clebsch ($\Beq=\gradclebscha\times\gradpolflux$) and axisymmetric ($\Beq=\polcurfunc\gradtorangle+\gradtorangle\times\gradpolflux$) representations of the equilibrium magnetic field, one can write an expression for $\clebscha$:
\begin{equation}
	\clebscha = \torangle - \int_{0}^{\polangle} d\dummy{\polangle}\locsafety,
\end{equation}
where $\locsafety(\minrad,\polangle)\equiv\Beq\cdot\gradtorangle/\Beq\cdot\gradpolangle$ is the local safety factor. It follows that $\gradclebscha$ contains the local magnetic shear $\locshear\equiv\minrad(\log\locsafety)'$ and can be written as
\begin{equation}
	\gradclebscha = \gradtorangle - \locsafety\gradpolangle - \gradminrad\int_{0}^{\polangle}d\dummy{\polangle}\locsafety',
\end{equation}
where $'$ indicates an $\minrad$-derivative. As shown in Figure \ref{fig:locshear}, the local magnetic shear affects the toroidal drive by shearing turbulent structures towards or away from the major radial direction in which they are most strongly driven \cite{antonsen1996,kessel1994}. To obtain expressions for the local magnetic shear as well as the other geometric coefficients appearing in the gyrokinetic system of equations, one can specify $\majrad(\minrad,\polangle)$ and $\heightZ(\minrad,\polangle)$ for a given flux-surface. A typical parametrization, and indeed the one we use here, is the Miller parametrization \cite{miller1998} which uses nine parameters to specify the poloidal cross section:
\begin{equation}
	\majrad(\minrad,\polangle) = \majradzero(\minrad) + \minrad\cos\big(\polangle + \arcsin\tri(\minrad) \sin\polangle\big)
\end{equation}
\begin{equation}
	\heightZ(\minrad,\polangle) = \elo(\minrad)\minrad\sin\polangle,
\end{equation}
where $\tri$ is the triangularity and $\elo$ is the elongation. Only first derivatives of $\minrad$-dependent quantities are specified, and the Grad-Shafranov equation  \cite{grad1958,shafranov1966} is used to ensure that the equilibrium locally satisfies MHD force balance. Consequently, $\betaprim\equiv\beta(\log\pressure)'$ must also be specified. More details on how to obtain expressions for $\locshear$ and $\magkperp^2$ are provided in Appendix \ref{app:geometry}.

\section{Results}
\label{sec:results}
\begin{figure}[htb!]
	\centering
	\includegraphics[width=0.8\linewidth]{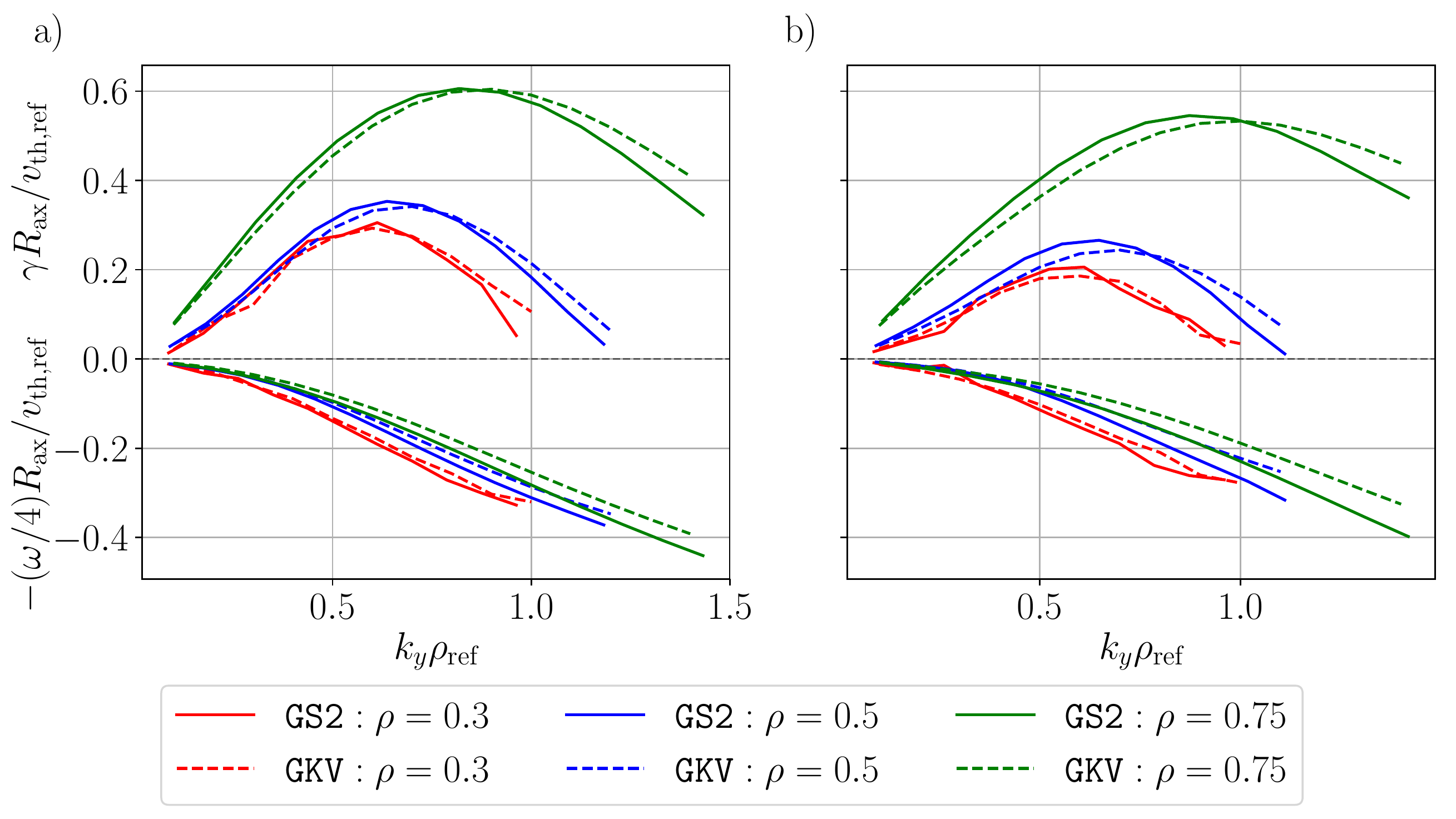}
	\caption{\label{fig:bench}Benchmark showing normalized real-frequency $\realfreq$ and growth-rate $\growthrate$ spectra at three different radial positions for the a) low-$\beta$ and b) high-$\beta$ equilibria. The \gstwo\ variables have been renormalized to their \gkv\ equivalents: $\majrad_\mathrm{ax}$ is the major radius of the magnetic axis, $\vthref$ is the reference thermal velocity and $\larmref$ is the reference Larmor radius.}
\end{figure}

\begin{figure}[htb!]
	\centering
	\includegraphics[width=0.6\linewidth]{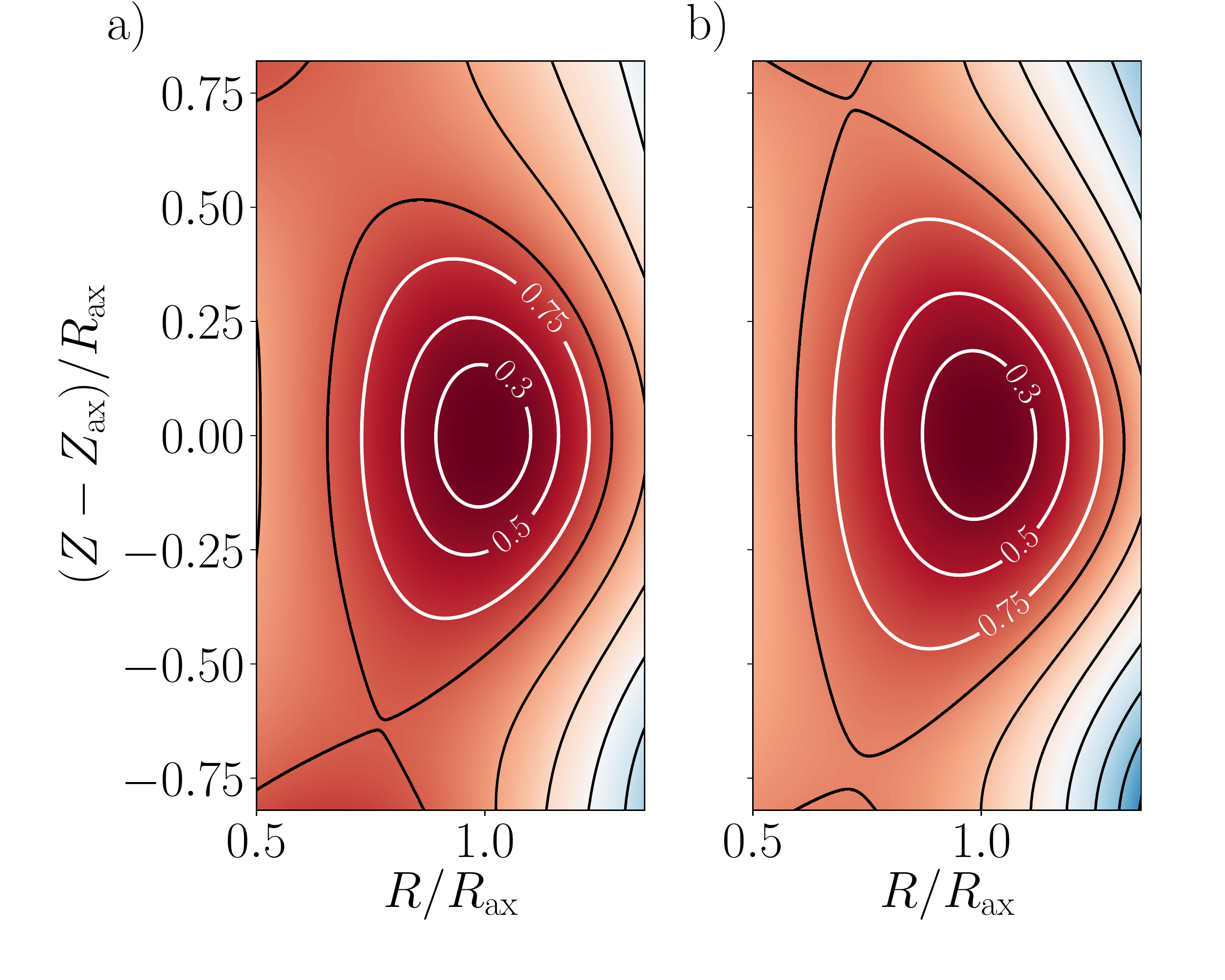}
	\caption{\label{fig:eqbm}Surfaces of constant magnetic flux for the a) low- and b) high-$\beta$ equilibria. White contours, labelled by their values of $\normtorflux$, denote surfaces used in the simulations.}
\end{figure}

\begin{figure}[htb!]
	\centering
	\includegraphics[width=\linewidth]{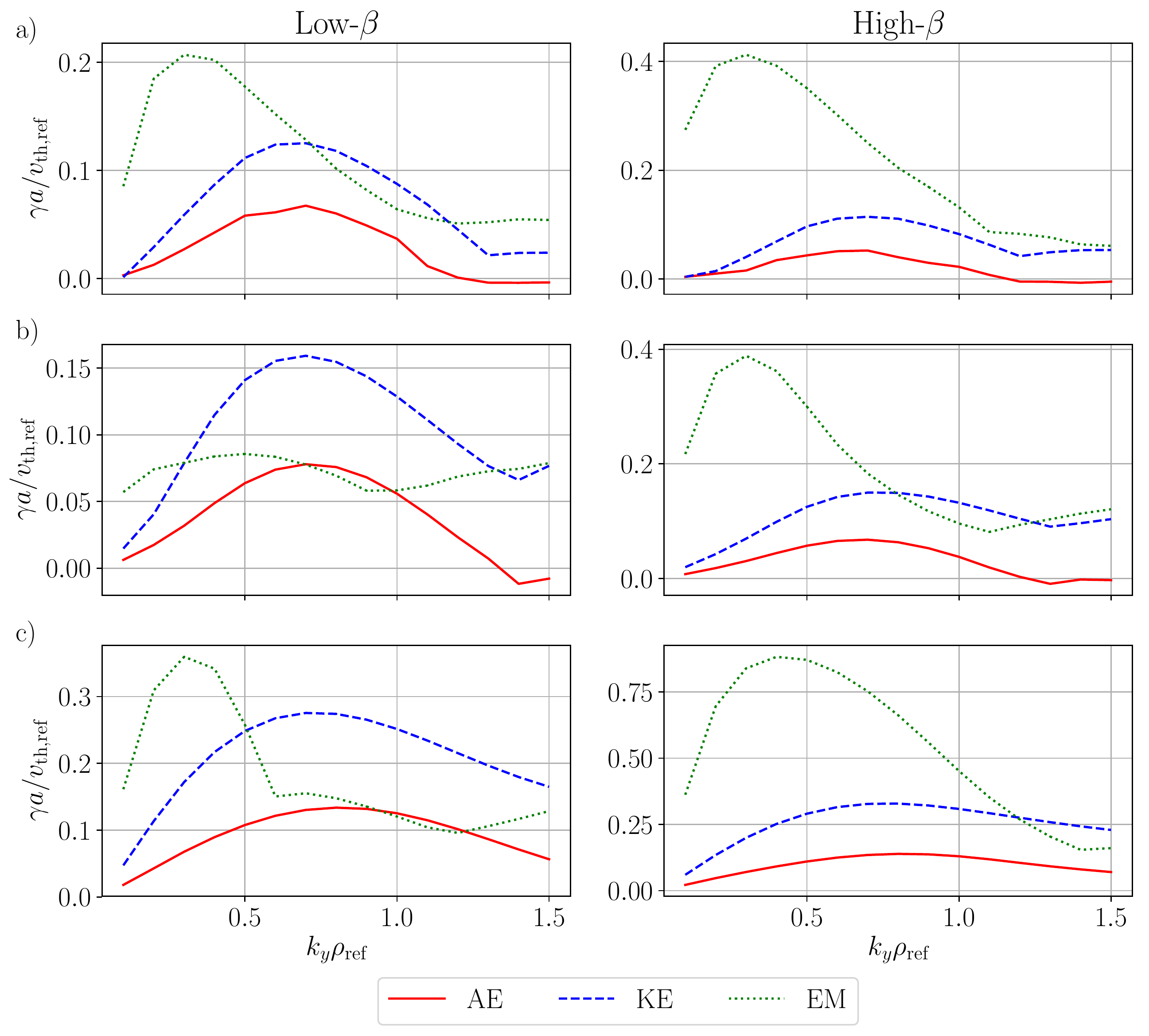}
	\caption{\label{fig:aekeem}Adding kinetic electrons and electromagnetic perturbations to the two equilibria at a) $\normtorflux=0.3$, b) $\normtorflux=0.5$ and c) $\normtorflux=0.75$. AE refers to electrostatic with adiabatic electrons, KE to electrostatic with kinetic electrons, whilst EM includes both kinetic electrons and electromagnetic fluctuations and $\macrolength$ is the half-diameter of the last closed flux-surface at the midplane.}
\end{figure}

Following the work of Nakata \cite{nakata2014}, we present the results of a benchmark between \gstwo\ and the analogous code \gkv, in Figure \ref{fig:bench}. The benchmark was performed at three radial ($\normtorflux$, the normalized toroidal magnetic flux) positions with adiabatic electrons and a single ion species, using Miller parametrizations of two numerical equilibria: one low-$\beta$ $(1.5\%)$ and one high-$\beta$ $(3.7\%)$ \cite{gs2tools}. The flux surface shapes and equilibrium parameters are given in Figure \ref{fig:eqbm} and Table \ref{tab:eqpars}. There is reasonable agreement between the linear growth rates $\growthrate$ and real frequencies $\realfreq$ calculated by the two codes. There is a small discrepancy that was also observed in a previous benchmark between the two codes \cite{mikkelsen2014} -- this may be due to algorithmic differences that affect the numerical dissipation. Despite this, the agreement was sufficient that we extended the study for the two equilibria by adding kinetic electrons and then electromagnetic perturbations; these results are shown in Figure \ref{fig:aekeem}. The addition of kinetic electrons approximately doubles all linear growth rates and destabilizes trapped electron modes (TEM) for $k_y\larmi\gtrsim1$, where the normalized poloidal wavenumber $k_y$ is related to $k_{\clebscha}$ via $k_y = k_{\clebscha}\Bref/\polflux'$. Adding full $\Bfluc$ perturbations destabilizes kinetic ballooning modes (KBMs) for both equilibria, though they are more potent in the high-$\beta$ case. These modes are identified as KBMs due a step change in the real frequency. Studies were also performed including \textit{either} $\Bflucpar$ \textit{or} $\Aflucpar$. With only $\Bflucpar$ compressive magnetic fluctuations, the results were almost identical to the electrostatic simulations. With only field line bending ($\Aflucpar$), a KBM-like mode appeared at low $\ky$, but with a growth rate approximately 50\% smaller than the full $\Bfluc$ fluctuations. In both equilibria, the electromagnetic effects stabilize ITG.

\Table{\label{tab:eqpars}Equilibrium parameters for the two equilibria. $\invasprat\equiv\minrad/\majradzero$ is the inverse-aspect ratio, $\shat$ is the magnetic shear, $\shift\equiv\majradzero'$ is the Shafranov shift and $\alpha\equiv-\betaprim\safety^2\majradzero$ where $\safety$ is the safety factor. $\tempe=\tempi$ for both equilibria.}
\br
&$\beta$&&\centre{3}{Low (1.5\%)}&\centre{3}{High (3.7\%)}\\
\mr
&$\rho$&&$0.3$&$0.5$&$0.75$& $0.3$&$0.5$&$0.75$\\
&$\invasprat$&&$0.10$&$0.17$&$0.25$&$0.12$&$0.21$&$0.30$\\
&$\safety$ && 1.85 & 2.02 &  2.66 & 1.37 & 1.55 & 2.23\\
&$\shat$&&$0.09$&$0.34$&$1.44$&$0.12$&$0.48$&$1.96$\\
&$\alpha$&&$0.48$&$0.83$&$1.19$&$0.62$&$1.13$&$1.66$\\
&$-\shift$&&$0.07$&$0.11$&$0.17$&$0.08$&$0.14$&$0.21$\\
&$\kappa$&&$1.50$&$1.52$&$1.58$&$1.49$&$1.52$&$1.61$\\
&$\kappa'$&&$0.04$&$0.10$&$0.33$&$0.05$&$0.14$&$0.47$\\
&$\delta$&&$0.08$&$0.14$&$0.23$&$0.10$&$0.17$&$0.29$\\
&$\delta'$&&$0.26$&$0.29$&$0.48$&$0.31$&$0.36$&$0.66$\\
&$(\log\temps)'$&&$2.33$&$2.42$&$2.72$&$2.69$&$2.84$&$3.33$\\
&$(\log\denss)'$&&$0.78$&$0.81$&$0.91$&$0.90$&$0.95$&$1.11$\\
\br
\end{tabular}
\end{indented}
\end{table}

\subsection{Electrostatic, linear shaping scans}

\begin{figure}[htb!]
	\centering
	\includegraphics[width=\linewidth]{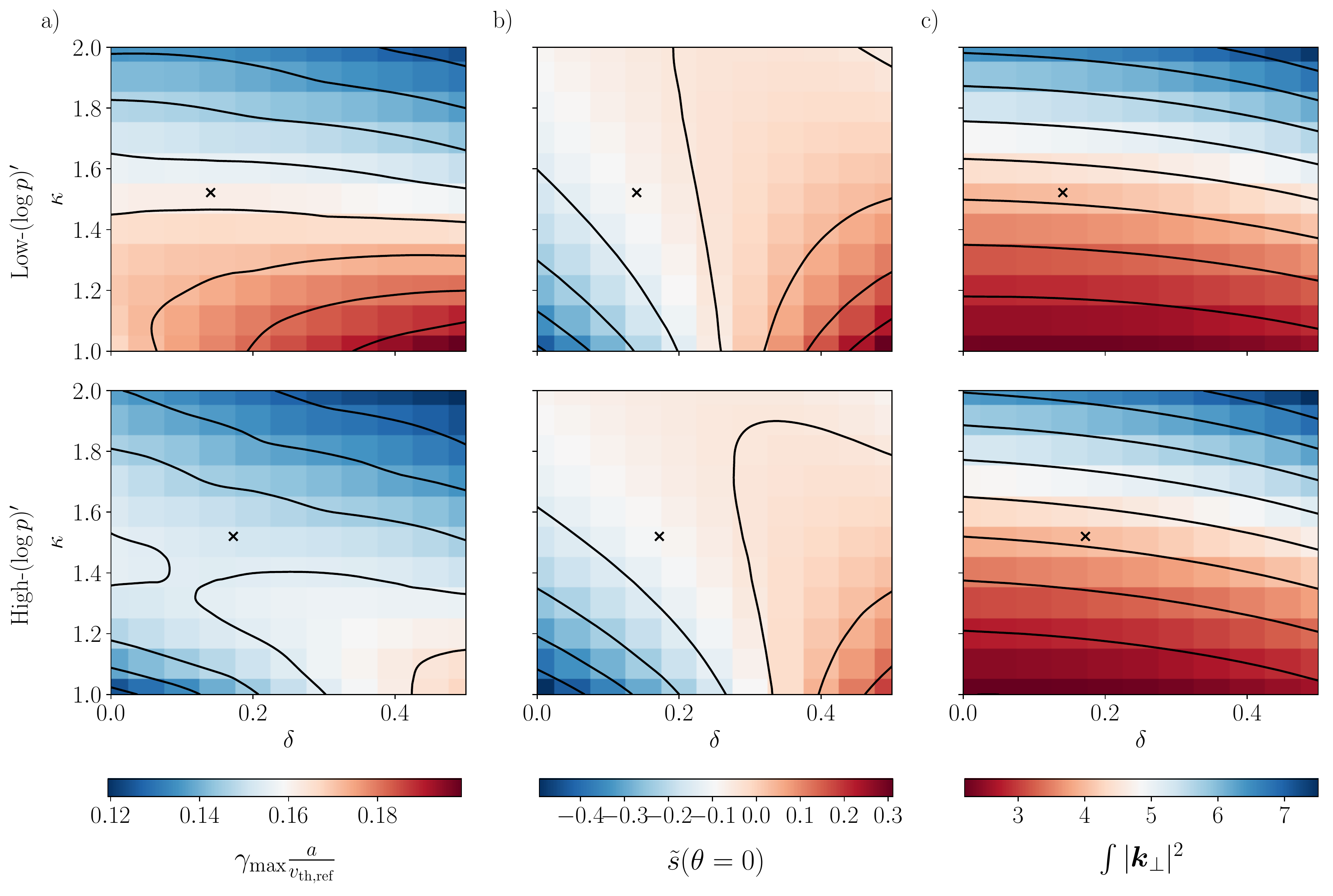}
	\caption{\label{fig:sextych}Scans in triangularity and elongation for two equilibria, showing a) linear growth rates, b) local magnetic shear at the outboard midplane, c) integral of $\magkperp^2$ over $-\pi/2\leq\polangle\leq\pi/2$. The black crosses indicate the nominal shapes. We note that the differences in local shear and $\int\magkperp^2$ between the two equilibria are small. This is because whilst the low-$\normpprim$ equilibrium has a normalized pressure gradient almost three times smaller than the high-$\normpprim$ one, the increased $\safety$ and $\majradzero$ result in a comparatively similar value of $\alpha$, which governs the overall effect of pressure gradient on the geometrical coefficients. The colours have been chosen such that blue indicates increased stability.}
\end{figure}

To determine the flux-surface shape that minimizes turbulent heat flux, electrostatic scans were performed in the triangularity and elongation Miller parameters for the $\normtorflux=0.5$ equilibria. All other parameters were held fixed; simultaneously scaling $\trir$ and $\elor$ with $\tri$ and $\elo$ was deemed unnecessary as their contributions are smaller by $\invasprat\ll1$. These electrostatic simulations are equivalent to low-$\beta$ electromagnetic simulations with $(\log\pressure)'$ modified to maintain $\betaprim\equiv\beta\normpprim$. As such, until electromagnetic effects are included, we refer to the equilibria as low- and high-$\normpprim$. The results of the scan are shown in Figure \ref{fig:sextych}a). For the low-$\normpprim$ equilibrium, increasing the elongation has an almost universal stabilizing effect, whereas the triangularity is destabilizing at low elongation and stabilizing at high elongation. The result is that maximal shaping minimizes the linear ITG instability. The maximum growth rates for the high-$\normpprim$ equilibrium are, for all $\lbrace\tri,\elo\rbrace$, less than for the low-$\normpprim$ equilibrium. Additionally, a moderate increase in elongation is destabilizing even up to $\tri\sim0.2$, with the result that circular flux surfaces are as stable as those with maximal shaping.

\begin{figure}[htb!]
	\centering
	\includegraphics[width=\linewidth]{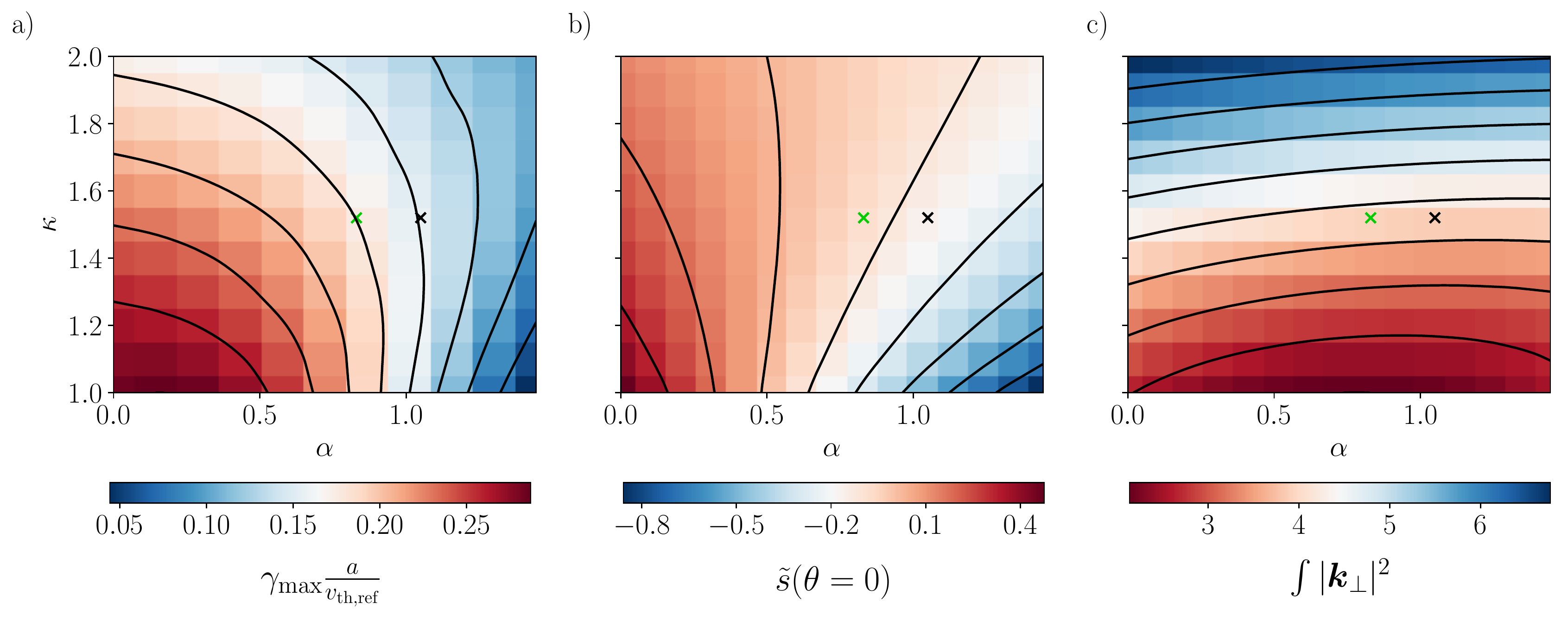}
	\caption{\label{fig:triptych}Electrostatic scans in $\alpha$
		and elongation for the high-$\normpprim$ equilibrium at nominal triangularity ($\tri=0.17$), showing a) linear growth rates maximised over $\ky$, b) local magnetic shear at the outboard midplane, c) integral of $\magkperp^2$ over $-\pi/2\leq\polangle\leq\pi/2$. The green and black crosses indicate the nominal parameters for the low- and high-$\beta$ equilibria, respectively. The scan in $\alpha$ was performed at fixed $\safety$ and $\majradzero$, so it corresponds to a scan in $\normpprim$. The nominal $\braced{\alpha}{\elo}$ value for the low-$\normpprim$ equilibrium is also shown by a green cross; a similar scan with the low-$\normpprim$ equilibrium parameters gives qualitatively similar results.}
\end{figure}

Using Figures \ref{fig:sextych}b) and c) we show that the trends in Figure \ref{fig:sextych}a) can be explained via a combination of local magnetic shear and FLR stabilization. To quantify the effect of local magnetic shear, we use its value at the outboard midplane where the ballooning modes are driven most strongly. Meanwhile, we quantify the FLR stabilization via the integral $\int_{-\pi/2}^{\pi/2}\magkperp^2\d\polangle$, which gives a measure of how much the mode is restricted over the bad-curvature region which extends approximately from $-\pi/2\leq\polangle\leq\pi/2$. For $\elo\simeq1$, the local magnetic shear is rapidly made less negative by increasing triangularity. In contrast, $\magkperp$ changes relatively slowly with triangularity, so the overall effect is a strong destabilization. At larger $\elo\simeq2$, the local shear is insensitive to changes in triangularity, so the increased $\magkperp^2$ gives a net stabilization with $\tri$. The effect of elongation on $\locshear$ changes with triangularity and $\alpha$. At large $\tri$, increased elongation makes the local shear more negative whilst also increasing $\magkperp^2$; these two effects work in tandem to provide strong stabilization. In Figure \ref{fig:triptych} we show that for $\tri\simeq0.2$, increased elongation reduces the local shear for small $\alpha\lesssim0.5$, but increases it for larger $\alpha\gtrsim0.5$. Both equilibria have sufficient $\alpha$ for increased elongation to increase $\locshear$, although the high-$\normpprim$ equilibria has higher $\alpha$ and so $\locshear$ is sufficiently sensitive to changes in $\elo$ that it outcompetes the increased FLR stabilization afforded by higher $\elo$. In contrast, the low-$\normpprim$ equilibrium, with lower $\alpha$, is dominated by the stabilizing effect of $\kperp^2$-stabilization. In Appendix \ref{app:geometry}, a simplified analytical model is developed; this gives qualitatively identical results to what is observed in Figure 7.

\subsection{Electrostatic, nonlinear shaping scans}
\begin{figure}[htb!]
	\centering
	\includegraphics[width=0.95\linewidth]{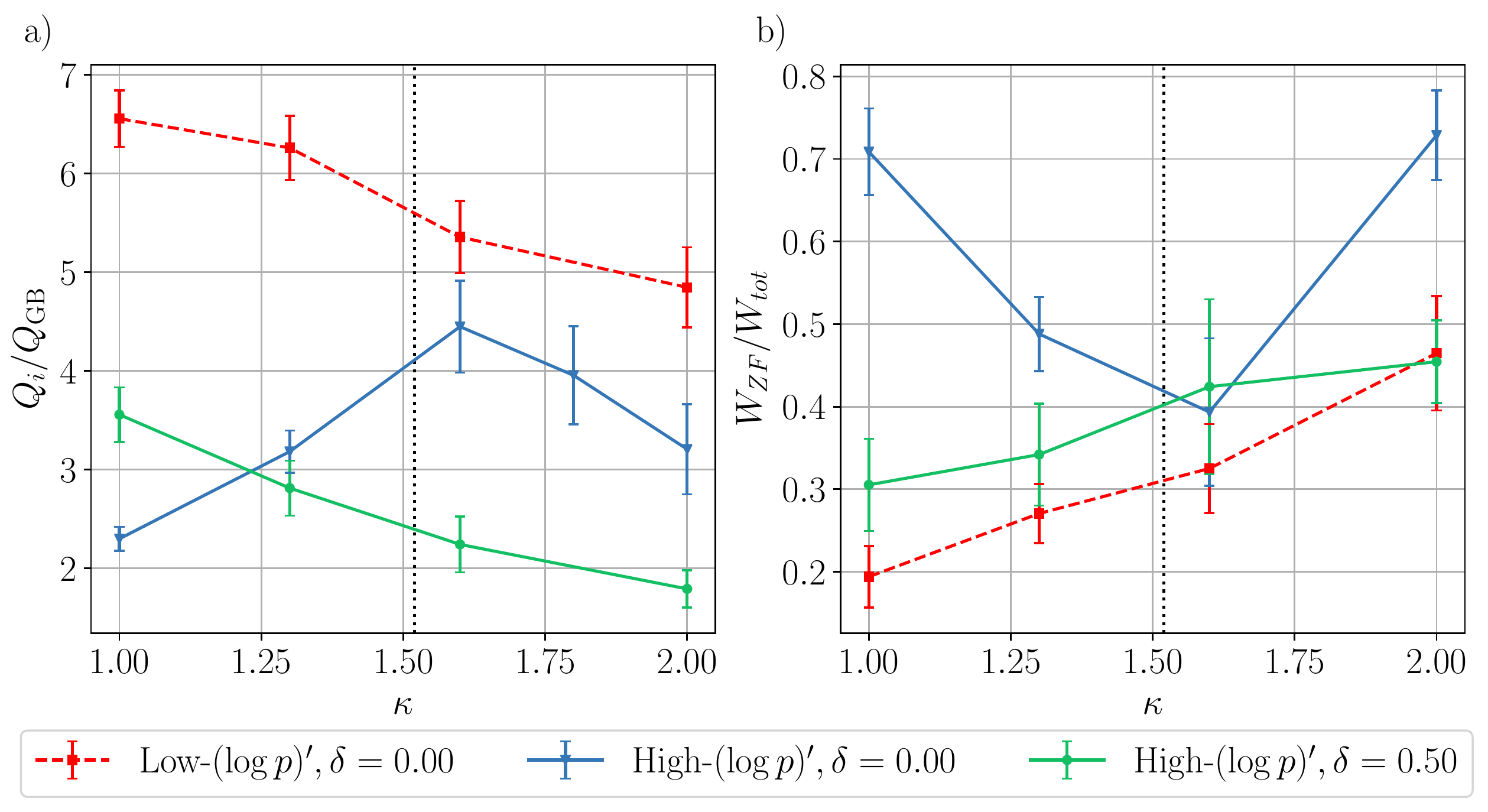}
	\caption{\label{fig:nlkap}a) Ion heat flux and b) fraction of zonal energy vs elongation for each equilibrium. The dotted line indicates the nominal elongation.}
\end{figure}

We also present non-linear electrostatic scans in elongation for both equilibria in Figure \ref{fig:nlkap}. All simulations were carried out with $32$ parallel grid points, 22 $k_y$ values, 256 $k_x$ values, with $\Delta k_x\simeq\Delta k_y = 0.047$. The polar velocity space grid contained 16 radial energy grid points, 20 passing pitch angles and up to 33 trapped pitch angles \cite{barnes2010, kotsch1995}. The trends observed in heat flux are similar to those of the linear growth rates: at $\tri=0$, increased elongation monotonically stabilizes the low-$\normpprim$ equilibrium whilst the high-$\normpprim$ case is destabilized up to $\elo\simeq1.6$. We note that the different $\elo$-position of the maximum (c.f. $\elo\simeq1.2$ in the linear results) is the main discrepancy between the linear and non-linear results. At $\tri=0.5$, increased elongation monotonically stabilizes the high-$\normpprim$ equilibrium. These results again suggest that maximal shaping maximizes performance for the low-$\beta$ plasma with moderate pressure gradient. However, in a low-$\beta$ plasma with steep pressure gradient, similar transport levels could be achieved with circular flux-surfaces as with maximal shaping.

\paragraph{}The fraction of energy in the zonal flow is also plotted in Figure \ref{fig:nlkap}b). To calculate this, we take the ratio of the following proxies for the zonal flow energy:
\[ W_{\mathrm{ZF}} \propto \bigg\langle\sum_{\kx} \big(1-\Gamma_{\kx,\ky=0}(\polangle)\big)\bigabs{\flucpot_{\kx,\ky=0}(\polangle)}^2\bigg\rangle_\polangle,\]
and the total energy contained in both turbulence and zonal flows:
\[ W_{\mathrm{tot}} \propto \bigg\langle\sum_{\kx,\ky} \big(1-\Gamma_{\kx,\ky}(\polangle)\big)\bigabs{\flucpot_{\kx,\ky}(\polangle)}^2\bigg\rangle_\polangle\]
where $\Gamma_{\kx,\ky}(\polangle)$ denotes the gamma function evaluated at $k_\perp(\kx,\ky,\polangle)\larmref$ and $\langle\hspace{0.5mm}\dots\rangle_\polangle$ indicates an average over poloidal angle \cite{nakata2017}. For the low-$\normpprim$ equilibrium at $\tri=0$ and the high-$\normpprim$ at $\tri=0.5$ the zonal energy fraction increases monotonically with elongation, reflecting the decreased transport. Similarly, for the high-$\normpprim$, ${\tri} = 0$ simulation the minimum zonal energy fraction coincides with the maximum heat flux. These results suggest that the effect of elongation on turbulent transport is closely linked to its effect on the relative strength of the zonal flow. 




\subsection{Electromagnetic shaping scans}

\begin{figure}[htb!]
	\centering
	\includegraphics[height=7cm]{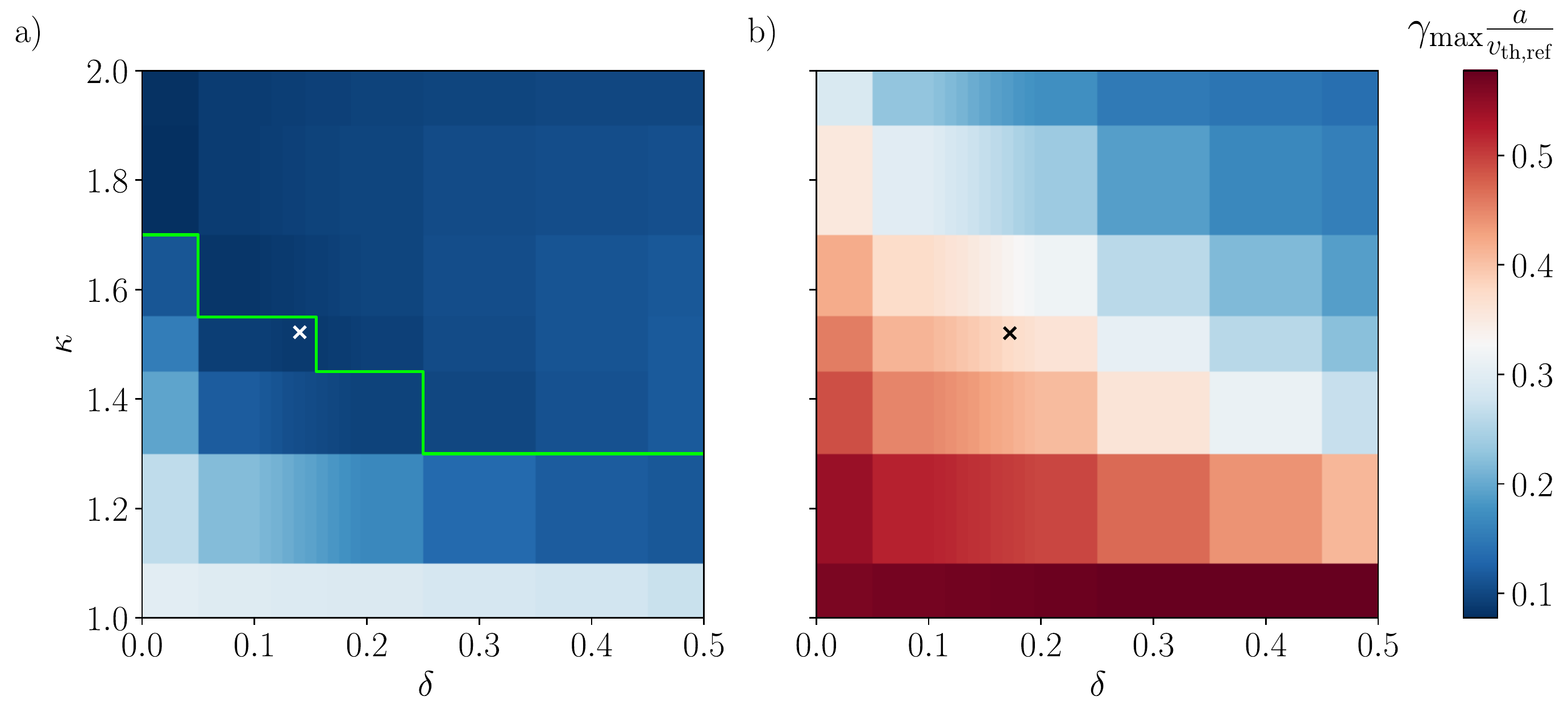}
	\caption{\label{fig:emtrikap}Electromagnetic scans in triangularity and elongation for the a) low-$\beta$ and b) high-$\beta$ equilibria, showing maximum linear growth rates. Below the green line, the fastest growing mode is the KBM. All modes in the high-$\beta$ case are KBMs. The crosses indicate the nominal shape.} 
\end{figure}

In Figure \ref{fig:aekeem} electromagnetic effects are observed to have a significant effect on the linear growth-rate spectra. We therefore include electromagnetic effects in the shaping studies. Figure \ref{fig:emtrikap} shows the dependence of the maximum linear growth rates on  triangularity and elongation for both equilibria. 
In the high-$\beta$ equilibrium, a KBM is excited for all $\lbrace\tri,\elo\rbrace$, and the growth rates are everywhere higher than the electrostatic case. Furthermore, the effect of plasma shaping becomes uniformly stabilizing. The nominal low-$\beta$ equilibrium is on the threshold of KBM stability, as indicated by a step change in the wavenumber and real frequency of the fastest growing mode either side of the green line in Figure \ref{fig:emtrikap}a). In contrast to the electrostatic results, the low-$\beta$ equilibrium has everywhere lower linear growth rates than the high-$\beta$ case, and triangularity is no longer destabilizing at low elongation.

\paragraph{}These results can be explained by an increased sensitivity of the KBM to FLR stabilization, compared to the electrostatic ITG. The local magnetic shear is most sensitive to triangularity at $\elo=1$, and only here is it able to compete with the $\magkperp^2$ stabilization, giving the almost constant linear growth rates at $\elo=1$. For all other elongations, the KBM is not sufficiently sensitive to $\locshear$ and is thus stabilized by $\magkperp^2$ that increases with both $\tri$ and $\elo$. As mentioned, the KBM is the dominant mode for all $\lbrace\tri,\elo\rbrace$ at high $\beta$, but for the low-$\beta$ case it becomes sub-dominant for strong shaping. 

\begin{figure}[htb!]
	\centering
	\includegraphics[width=0.8\linewidth]{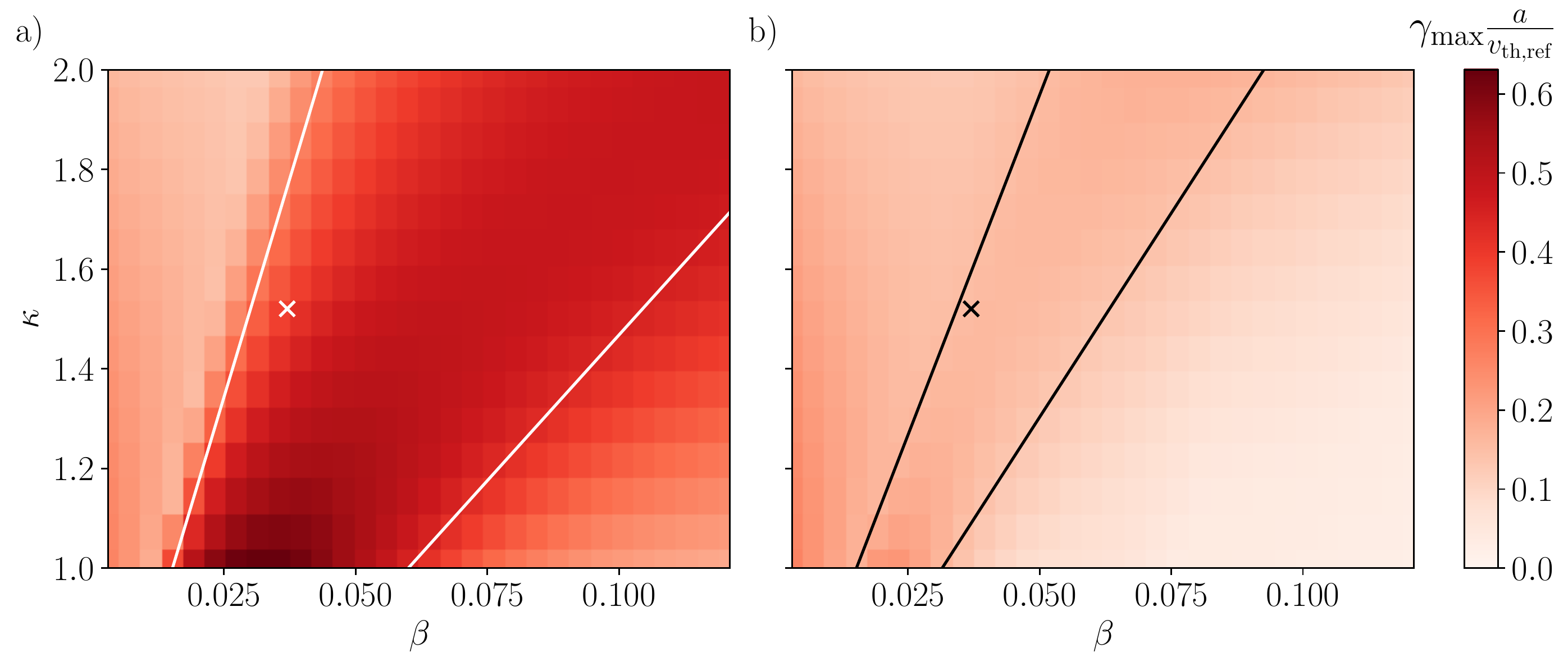}
	\caption{\label{fig:embetakap}Electromagnetic scans in $\beta$ and elongation for the high-$\beta$ equilibrium, with $(\log\pressure)'$ held fixed at a) nominal and b) 1.5 times nominal. The crosses indicates the nominal $\lbrace\beta,\elo\rbrace$, whilst the lines are shown to guide the eye to approximate bounds for the region of KBM instability (they have not been determined via MHD calculations). We reiterate that increased $\beta$ leads to more negative local magnetic shear.}
\end{figure}

\paragraph{}In Figure \ref{fig:embetakap} we show scans in $\beta$ at fixed $(\log\pressure)'$. As $\beta$ increases, $\locshear(\polangle=0)$ is reduced via the corresponding increase in $\alpha\propto\beta$. Two such scans are shown, at nominal and 1.5 times nominal $\normpprim$.
At low $\beta$, we observe the $\beta$-stabilization of ITG modes, and note that the KBM threshold $\beta$ increases with elongation due to FLR stabilization, consistent with Figure \ref{fig:emtrikap}. We also observe the persistence of KBMs (albeit with significantly reduced linear growth rates) into the region of second MHD stability, as previously reported in \cite{hirose1994}, for sufficiently high $\beta$ approaching $10\%$. The reduction in growth rates in this region can be attributed to the increasing local-shear-stabilization due to larger $\alpha\propto\beta$. As $\elo$ is increased in the region of second stability, the local-shear stabilization is reduced rapidly enough that it outcompetes FLR stabilization. This has the effect of removing the second stability afforded by increased $\beta$. 

\paragraph{}To summarize, the differences in the results obtained from the nominal- and increased-$\normpprim$ scans are as follows: the growth rates are everywhere lower in the increased-$\normpprim$ case, and the KBM reacts more strongly to increased $\normpprim$ compared to the ITG. This is because the absolute change in $\alpha$ scales linearly with $\beta$, so the local magnetic shear is less reduced at low-$\beta$. In the increased-$\normpprim$ case, the second stability region is accessed at lower $\beta$. This in turn lowers the threshold $\beta$ at which $\elo$ becomes destabilizing. This suggests that to minimize turbulent transport in a high-performance plasma with steep pressure gradients and high $\beta$, the flux surface elongation should be minimized. 

\section{Summary}
\label{sec:summary}
\paragraph{}In this work we have performed a successful benchmark between the local $\delta f$-gyrokinetic codes $\gstwo$ and $\gkv$ for two equilibria with significant shaping effects. We have also performed, with the inclusion of electrons as a kinetic species, linear scans in the flux-surface shaping parameters $\tri$ and $\elo$ to study the effect on turbulent transport. We find that for low-$\beta$ equilibria, triangularity is (de)stabilizing at (low) high elongation, and vice versa. We also find the novel result that elongations of $\elo\sim1.5$ can be destabilizing in regions of steep normalized pressure gradient. This trend in linear growth rate is also observed in the non-linear heat-flux. We explain these results as a competition between the effects of shaping on the local magnetic shear and on FLR stabilization. Electromagnetic scans were also performed, and KBMs were destabilized for both equilibria. Our results show that for the nominal equilibrium parameters, the KBM is stabilized monotonically by increased shaping; we infer from this that for the KBM, FLR stabilization is stronger relative to the local magnetic shear. Scans in $\lbrace\beta,\elo\rbrace$ show that at sufficiently high $\beta$, the local magnetic shear becomes sensitive enough to $\elo$ that it can compete with FLR stabilization and compromise second-stability. The threshold $\beta$ at which this occurs decreases with increased normalized pressure gradient. Therefore, we suggest that for a high-performance plasma, i.e. one with high-$\beta$ and steep pressure gradients, the turbulent outwards radial fluxes may be minimized by reducing elongation as much as possible. Conversely, for a moderate $\beta$ plasma, we would suggest maximal elongation to FLR-stabilize the KBMs. 

\section*{Acknowledgements}
This work has been carried out within the framework of the EUROfusion Consortium and has received funding from the Euratom research and training programme 2014-2018 and 2019-2020 under grant agreement No 633053. The views and opinions expressed herein do not necessarily reflect those of the European Commission.

\begin{appendices}
	\section{Calculation of the local magnetic shear and perpendicular wavenumber}\label{app:geometry}
	
	In this section we flesh out the details of how to determine the effect of shaping parameters on stability. As an example, we show an analytic calculation of the effect of elongation to leading order in $\invasprat\ll1$. As discussed in Section \ref{sec:model}, this involves determining the local magnetic shear and the perpendicular wavenumber.
	
	\paragraph{}The local safety factor $\locsafety$ is the ratio of the toroidal to poloidal component of the magnetic field:
	\begin{equation}
		\locsafety \equiv \frac{\magBeq\cdot\gradtorangle}{\magBeq\cdot\gradpolangle} = \frac{\polcurfunc\jacr}{\majrad^2\polflux'}
	\end{equation}
	where we used the axisymmetric form of the magnetic field:
	\begin{equation}
		\Beq=\polcurfunc\gradtorangle + \gradtorangle\times\gradpolflux
	\end{equation}
	and defined the Jacobian $\jac_X$ for the transformation between $\lbrace\majrad,\torangle,\heightZ\rbrace\to\lbrace X,\polangle,\torangle\rbrace$ where $X$ is any flux-surface label. This also allows us to define $\polflux'$ in terms of the safety factor $\safety$:
	\begin{equation}
		\polflux' = \frac{\polcurfunc}{2\pi\safety}\int_{0}^{2\pi}\d\polangle\frac{\jacr}{\majrad^2}\label{eq:polflux'}
	\end{equation}
	 Therefore, the local magnetic shear is
	\begin{equation}
		\locshear \equiv \minrad\frac{\locsafety'}{\locsafety} = r\bigg( \frac{\polcurfunc'}{\polcurfunc} + \frac{\jacprim}{\jacr} - 2\frac{\majrad'}{\majrad}\bigg). 
	\end{equation}
	where $\jacprim \equiv\jacr' - \jacr\polflux''/\polflux'$. To evaluate this, we use the Grad-Shafranov equation:
	\begin{equation}
		\majrad^2\nabla\cdot\bigg( \frac{\gradpolflux}{\majrad^2} \bigg) = -\frac{\polcurfunc\polcurfunc' + 4\pi\majrad^2\pressure'}{\polflux'}
	\end{equation}
	By expanding the divergence and substituting explicit forms for $\gradminrad$ and $\gradpolangle$ using a prescribed $\majrad(\minrad,\polangle)$ and $\heightZ(\minrad,\polangle)$, we arrive at the following expression for $\jacprim/\jacr$:
	\begin{equation}
	\frac{\jacprim}{\jacr}= \frac{\majrad^2}{\jacr^2\abs{\gradminrad}^2}\Bigg[2\bigg( \pdshort{\majrad{'}}{\polangle}\pdshort{\majrad}{\polangle} + \pdshort{\heightZ{'}}{\polangle}\pdshort{\heightZ}{\polangle}\bigg) - \jacr\pd{}{\polangle}\bigg( \frac{1}{\jacr}\bigg[ \majrad'\pdshort{\majrad}{\polangle} + \heightZ'\pdshort{\heightZ}{\polangle} \bigg]\bigg) \Bigg] + \frac{\polcurfunc\polcurfunc'+4\pi\majrad^2\pressure'}{\abs{\gradpolflux}^2} \label{eq:jacprim},
	\end{equation}  
	where superscript $\polangle$ indicates a $\polangle$-derivative.
	To get an expression for $\polcurfunc'/\polcurfunc$, we take the $\minrad$-derivative of Equation (\ref{eq:polflux'}) and move all terms under the integral:
\begin{equation}0 = \int_{0}^{2\pi}\d{\polangle}\Bigg[\frac{\jacprim}{\majrad^2} + \frac{\jacr}{\majrad^2}\bigg( \frac{\polcurfunc'}{\polcurfunc} -\frac{\safety'}{\safety} - \frac{2\majrad'}{\majrad} \bigg)\Bigg].
\end{equation}
	This can be used in conjunction with Equation (\ref{eq:jacprim}) integrated over $\polangle$ to give an expression for $\polcurfunc'$:
	\begin{equation}\frac{\polcurfunc'}{\polcurfunc}\int_{0}^{2\pi}\d\polangle\frac{\jacr}{\majrad^2}\bigg( 1+\frac{\polcurfunc^2}{\abs{\nabla \polflux}^2} \bigg) = \int_{0}^{2\pi}\d\polangle\frac{\jacr}{\majrad^2}\bigg(\frac{\safety'}{\safety}+\frac{2\majrad'}{\majrad} - \frac{4\pi\majrad^2\pressure'  }{ \abs{\gradpolflux}^2 }\bigg)\nonumber \end{equation}
	\begin{equation}
		+\int_{0}^{2\pi}\frac{\d\polangle}{\abs{\gradminrad}^2}\Bigg( \pd{}{\polangle}\bigg(\frac{1}{\jacr}\bigg[ \majrad'\pdshort{\majrad}{\polangle}+\heightZ'\pdshort{\heightZ}{\polangle} \bigg] \bigg) - \frac{2}{\jacr}\bigg( \pdshort{\majrad{'}}{\polangle}\pdshort{\majrad}{\polangle} + \pdshort{\heightZ{'}}{\polangle}\pdshort{\heightZ}{\polangle} \bigg)\Bigg)\label{eq:Iprime}.
	\end{equation}
	With these expressions, one can generate expressions for $\locshear$ and thus $\magkperp^2$. 
	
	\subsection{Analytical expressions for concentric elliptical flux-surfaces}
	
	We next use a simplified Miller parametrization that includes only elongation to determine the effect of elongating a circular plasma:
	\begin{equation}
		\majrad(\minrad,\polangle) = \majradzero + \minrad\cos\polangle
	\end{equation}
	\begin{equation}
	\heightZ(\minrad,\polangle) = \elo\minrad\sin\polangle
	\end{equation}
	To leading order in inverse aspect ratio $\invasprat\ll1$, we proceed to determine $\locshear$ and $\magkperp^2$, beginning with the following quantities:
	\begin{equation}
		\jacr(\minrad,\polangle) = \majrad\big( \majrad'\pdshort{\heightZ}{\polangle} - \pdshort{R}{\polangle}\heightZ' \big) = \elo\minrad\majrad
	\end{equation}
	\begin{equation}\abs{\gradminrad}^2\equiv\frac{
		\majrad^2}{\jacr^2}\bigg( \big(\pdshort{\majrad}{\polangle}\big)^2 + \big( \pdshort{\heightZ}{\polangle} \big)^2 \bigg) = \frac{1+(\elo^2-1)\cos^2\polangle}{\elo^2}
	\end{equation}
	\begin{equation}\abs{\gradpolangle}^2\equiv\frac{
		\majrad^2}{\jacr^2}\bigg( \big({\majrad}'\big)^2 + \big( {\heightZ}' \big)^2 \bigg) = \frac{1+(\elo^2-1)\sin^2\polangle}{\minrad^2\elo^2}
	\end{equation}
	\begin{equation}\gradminrad\cdot\gradpolangle\equiv-\frac{
		\majrad^2}{\jacr^2}\bigg( \majrad'\pdshort{\majrad}{\polangle} + \heightZ' \pdshort{\heightZ}{\polangle}  \bigg) =  -\frac{(\elo^2-1)\sin\polangle\cos\polangle}{\minrad\elo^2}
	\end{equation}
	
	\begin{equation}
		\polflux' = \frac{\polcurfunc\elo\invasprat}{\safety}+\ord{\polcurfunc\invasprat^3}
	\end{equation}
	To determine $\polcurfunc'/\polcurfunc$, we calculate each of the three integrals that appear in Equation \ref{eq:Iprime} separately:
	\begin{equation}
		\int_{0}^{2\pi}\d\polangle\frac{\jacr}{\majrad^2}\bigg( 1 + \frac{\polcurfunc^2}{\abs{\gradpolflux}^2} \bigg) = \frac{2\pi\safety^2}{\invasprat} + \ord{\epsilon}
	\end{equation}
	\begin{equation}
		\int_{0}^{2\pi}\d\polangle\frac{\jacr}{\majrad^2}\bigg( \frac{\safety'}{\safety} + \frac{2\majrad'}{\majrad} - \frac{4\pi\majrad^2\pressure'}{\abs{\gradpolflux}^2}\bigg) = \frac{2\pi}{\invasprat\majradzero}\bigg\lbrace \elo\invasprat\shat + \frac{\alpha}{2} \bigg\rbrace + \ord{\invasprat/\majradzero}
	\end{equation}
	\begin{equation}
		\int_{0}^{2\pi}\frac{\d\polangle}{\abs{\gradminrad}^2}\Bigg( \pd{}{\polangle}\bigg(\frac{1}{\jacr}\bigg[ \majrad'\pdshort{\majrad}{\polangle}+\heightZ'\pdshort{\heightZ}{\polangle} \bigg] \bigg) - \frac{2}{\jacr}\bigg( \pdshort{\majrad{'}}{\polangle}\pdshort{\majrad}{\polangle} + \pdshort{\heightZ{'}}{\polangle}\pdshort{\heightZ}{\polangle} \bigg)\Bigg) = -\frac{2\pi}{\majradzero}\big( {\elo^2+1} \big) + \ord{\invasprat/\majradzero}.
	\end{equation}
	Combining these, we find that
	\begin{equation}
		\frac{\polcurfunc'}{\polcurfunc}\frac{\safety^2}{\invasprat}\majradzero = \elo\bigg( \shat - \frac{\elo^2+1}{\elo} \bigg)+\frac{\alpha}{2}\frac{1}{\invasprat} + \ord{\invasprat},
	\end{equation}
	where we will find that the leading order piece cancels with a term in $\jacprim/\jacr$, so we retain the $\ord{1}$ terms. Using this, we find:
	\begin{equation}
		\frac{\jacprim}{\jacr} = \frac{1}{\invasprat\majradzero}\Bigg\lbrace \frac{\elo\shat - \alpha\cos\polangle }{1+(\elo^2-1)\cos^2\polangle}\Bigg\rbrace.\nonumber
	\end{equation}
	Comparing the size of the terms that comprise $\locshear$, the largest is $\jacprim/\jacr$. Then, to leading order, we find 
	\begin{equation}
		\locshear = \frac{\elo\shat - \alpha\cos\polangle }{1+(\elo^2-1)\cos^2\polangle}.
	\end{equation}
	Now we can use $\locshear$ to find $\gradclebscha$ and thus $\magkperp^2$:
	\begin{equation}
		\int_{0}^{\polangle}\d\dummy{\polangle}\locsafety' = \frac{\safety}{\minrad}\bigg[ \shat\atanterm - \alpha\Lambda \bigg],
	\end{equation}
	where
	\begin{equation}
		\Lambda(\polangle,\elo) \equiv \lnterm
	\end{equation}
	and
	\begin{equation}
	\atanterm(\polangle,\elo) \equiv \atantermdef.
	\end{equation}
	which gives:
	\begin{equation}
		\magkperp^2 \propto \big( \shat(\atanterm-\theta_0) - \alpha\Lambda \big)^2\big( 1+ (\elo^2-1)\cos^2\polangle \big) +1+ (\elo^2-1)\sin^2\polangle \nonumber
	\end{equation}\vspace{-6mm}
	\begin{equation}-2\big( \shat(\atanterm-\theta_0) - \alpha\Lambda \big)(\kappa^2-1)\sin\polangle\cos\polangle + \mathcal{O}(\invasprat)
	\end{equation}
	where $\theta_0\equiv k_x/(\shat k_y)$. These expressions reduce to the typical $\shat-\alpha$ results when $\elo=1$, since $\Lambda(\polangle,1)=\sin\polangle$ and $\varpolangle(\polangle,1) = \polangle$.
	\end{appendices}

\end{document}